# A Family of Likelihood Ascent Search Multiuser Detectors: Approach to Single-User Performance via Quasi-Large Random Sequence CDMA

Yi Sun[1]

*Abstract* – **Since Tse and Verdú proved that the global maximum likelihood (GML) detector achieves unit asymptotic multiuser efficiency (AME) in the limit of large random spreading (LRS) CDMA, no suboptimal detector has been found to achieve unit AME. In this letter, we obtain that the WSLAS detector with a linear per-bit complexity achieves unit AME in the LRS-CDMA with a channel load < ½ − 1/(4ln2) bits/s/Hz. For a practical system with any user number, a quasi LRS-CDMA is then proposed to approach the single-user performance in the high SNR regime**.

## I. INTRODUCTION

Tse and Verdú proved [1] that the NP-hard global maximum likelihood (GML) detector achieves unit asymptotic multiuser efficiency (AME) in the limit of large random spreading (LRS) CDMA, and thus achieving the single-user bit error rate (BER) in the high SNR regime. Since then, none of suboptimal detectors with a practical complexity has been proved to achieve the unit AME except that there exist nontrivial CDMA channels where the LAS detector can achieve unit AME regardless of user number as shown in the companion paper [6]. The belief propagation (BP) algorithm is recently applied to approach the GML detection in the LRS-CDMA [2]-[5]. However, in order to reduce the complexity of BP, which grows exponentially with the user number, to a practical complexity, Gaussian approximation is necessarily applied.

The contribution of this letter is twofold. First, it is proved that the WSLAS detectors[2] [6] achieve the unit AME in the LRS-CDMA limit with a channel load less than ½ − 1/(4ln2) bits/s/Hz. Alike the result of Tse and Verdú [1], the unit AME is obtained by taking zero noise limit first and then large system limit. In a looser limit condition that the noise power tends to zero and the system size tends to infinity at the same rate, it is further proved that a local maximum likelihood (LML) point is identical to the GML point. The results are applicable to all LML detectors with any neighborhood size [7]. Second, by a scheme of bit extending and multiplexing, a quasi-large random sequence (QLRS) CDMA is proposed for a practical CDMA system with any user number to approach the single-user performance. The scheme does not incur either increase of bandwidth or/and power or decrease of transmission rate. Simulation results show that the LAS detectors can approach the GML BER of LRS-CDMA limit [11] with all SNR and approach the single-user BER in high SNR when the total number of bits is greater than 500 and the channel load is not greater than 1 bit/s/Hz. Since the single-user bound is approached with the channel load as high as 1 bit/s/Hz, the QLRS-CDMA employing the LAS detectors has approached the bound that a perfect TDMA or FDMA can approach. The per-bit complexity of the LAS detectors demonstrated in all simulations is less than 0.5 times the bit number.

---

[1] Yi Sun is with the Department of Electrical Engineering at the City College of City University of New York, New York, NY 10031. Phone: (212)650-6621; E-mail: ysun@ee.ccny.cuny.edu. This paper was presented in part at the 41st Annual Conference on Information Science and Systems, Baltimore, Maryland, March 14-16, 2007.
[2] All WSLAS detectors are local maximum likelihood (LML) detectors with neighborhood size one.

## II. PROPOSED QLRS-CDMA

### A. LRS-CDMA channel

Consider a $K$-user bit-synchronous Gaussian CDMA channel. The bit period is $T_b$ and the chip period is $T_c$. Then the transmission rate for uncoded bits per user is $1/T_b$ bits per second, the channel bandwidth approximately equals $W = 1/T_c$ and the spectral spreading factor equals $N = T_b/T_c$. The chip matched filter (MF) at the receiver outputs

$$\mathbf{r} = \sum_{k=1}^{K} A_k \mathbf{s}_k b_k + \mathbf{m} = \mathbf{SAb} + \mathbf{m}. \qquad (1)$$

$\mathbf{S} = (\mathbf{s}_1, \ldots, \mathbf{s}_K)$ where the $N$-chip $\mathbf{s}_k$ is the $k$th user's spreading sequence with unit length $\|\mathbf{s}_k\| = 1$. $\mathbf{A} = \text{diag}(A_1, \ldots, A_K)$ where $A_k$ is the received signal amplitude of user $k$. $\mathbf{b} = (b_1, \ldots, b_K)^T$ is the vector of $K$ users' transmitted bits that independently equiprobably take on $\pm 1$'s. $\mathbf{m} \sim N(\mathbf{0}, \sigma^2 \mathbf{I})$ is a white Gaussian noise vector. The MF bank $\mathbf{S}$ outputs a sufficient statistic $\mathbf{y} = \mathbf{S}^T \mathbf{r} = \mathbf{RAb} + \mathbf{n}$ where $\mathbf{R} = \mathbf{S}^T \mathbf{S}$ and $\mathbf{n} = \mathbf{S}^T \mathbf{m} \sim N(\mathbf{0}, \sigma^2 \mathbf{R})$.

In the large random sequence (LRS) CDMA, the user number and the sequence dimension tend to infinity with their ratio kept a constant $\alpha = K/N \in (0, \infty)$. The type of random spreading sequence $\mathbf{s}_k = (s_{1k}, \ldots, s_{Nk})^T / \sqrt{N}$ where $s_{jk}$ independently equiprobably takes on $\pm 1$'s can be practically short that a user initially selects a sequence and uses it to spread its all bits, or be practically long that a user newly selects a sequence for each bit. The received amplitudes $A_k$ are fixed regardless of sequence selection and are bounded by $A' \leq A_k \leq A''$ as $K$ tends to infinity. The channel load is equal to $\alpha = K/(WT_b) = K/N$ bits/s/Hz. With fixed $\alpha$, the bandwidth share per user $W/K = 1/(\alpha T_b)$ is unchanged and therefore the bandwidth usage is fixed as $K$ tends to infinity.

Tse and Verdú has proved [1] that the AME of the GML detector converges almost surely to one in the LRS-CDMA for all $\alpha$; in contrast, the AME converges to zero and $1-\alpha$ for the MF and for the decorrelator and the MMSE, respectively.

The LRS-CDMA is theoretically significant but impractical since a practical system has a fairly small user number and bandwidth.

### B. QLRS-CDMA

For a practical CDMA system where $K$ and $W$ (or $N$) are *finite* and *fixed*, we propose to construct a quasi-large random sequence (QLRS) CDMA by a scheme of bit extending and multiplexing. User $k$ collects and simultaneously transmits $B_k$ bits $b_{kj}$, $j = 1, \ldots, B_k$, which are extended by a factor of $B$ to occupy $B$ bit periods of $BT_b$ seconds and spread by the $BN$-chip unit-length random sequences $\mathbf{s}_{kj} \in \{-1/\sqrt{BN}, 1/\sqrt{BN}\}^{BN}$. Though each user may multiplex a different number of bits $B_k$, all the extended bits have the same duration $BT_b$. During an extended bit period, the chip MF outputs

$$\mathbf{r} = \sum_{k=1}^{K} A_k \sum_{j=1}^{B_k} \mathbf{s}_{kj} b_{kj} + \mathbf{m} \qquad (2)$$

where $\mathbf{m} \sim N(\mathbf{0}, \sigma^2 \mathbf{I}_{BN})$. The bit multiplexing factors $B_k$ and extending factor $B$ tend to infinity and their ratios $\beta_k = B_k/B$ for $k = 1, \ldots, K$ are fixed. Then the channel load is fixed and equals

$$\alpha = \frac{1}{WBT_b} \sum_{k=1}^{K} B_k = \frac{1}{N} \sum_{k=1}^{K} \beta_k \qquad (3)$$

bits/s/Hz.

The constructed QLRS-CDMA is mathematically identical to an LRS-CDMA where there are $K$ classes of users, each class has the same user power, and the channel load equals $\alpha$. The scheme does not change the bandwidth $W$ and user powers. Moreover, when $\beta_k = 1$ for $k = 1, \ldots, K$, all users also have the same transmission rate $1/T_b$ as in the original system. However, as a by-product, adjusting $\beta_k$ can attain different transmission rates. Even for a system where spectrum is not spread with $N = 1$, the QLRS-CDMA can be still constructed.

The proposed QLRS-CDMA differs from the conventional CDMA schemes. First, in the conventional CDMA the sequences spread only in frequency and the sequence dimension is equal to the spectral spreading factor $N$. In the QLRS-CDMA, the sequences may spread not only in frequency but also in time. The total spreading factor $BN$ is equal to the temporal spreading factor $B$ times the spectral spreading factor $N$ with the sequence dimension increased by a fact of $B$. Second, the QLRS-CDMA is also different from the conventional multirate CDMA that aims at providing variable transmission rates for users with different qualities of service. Moreover, the bit multiplexing used in multicode CDMA does not improve the BER performance of the conventional MF detector and other interference-limited detectors. In particular, the scheme of bit extending and multiplexing with the same factor $B$ does not change the interference power to each user and so does not change the performance of an interference-limited detector. However, the QLRS-CDMA has the advantage that as $B$ increases, the channel eventually possesses the LML characteristic due to the long random sequences so that the LAS detectors can approach the GML detection in all SNR and approach the single-user performance in high SNR. Meanwhile, the QLRS-CDMA does not incur decrease of transmission rate though as a side product a variable transmission rate can be obtained by varying the number of multiplexed bits. Finally, unlike the conventional LRS-CDMA, the QLRS-CDMA can be practically implemented since the temporal spreading factor $B$ can be arbitrarily large.

## III. ACHIEVABILITY OF SINGLE-USER PERFORMANCE

*A. LML characteristic*

Since the QLRS-CDMA is identical to a particular LRS-CDMA with a finite number of user classes, the following analysis is focused on the LRS-CDMA and all the results are applicable to the QLRS-CDMA. The short sequences shall be considered and the results are in the almost sure convergence, which are applicable to long sequences in the deterministic convergence.

The notions of error vector, error weight, and indecomposable error vectors developed by Verdú [12] shall be employed. Let $l^{GML}(\varepsilon)$ be the hyperplane separating the transmitted signal $\mathbf{SAb}$ and the error signal $\mathbf{SA}(\mathbf{b}-2\varepsilon)$ optimally in terms of GML. Then the distance from $\mathbf{SAb}$ to the hyperplane $l^{GML}(\varepsilon)$ equals [12]

$$d^{\text{GML}}(\boldsymbol{\varepsilon}) = \sqrt{\boldsymbol{\varepsilon}^T \mathbf{H} \boldsymbol{\varepsilon}} \tag{4}$$

where $\mathbf{H} = \mathbf{ARA}$. For the indecomposable error vector $\boldsymbol{\varepsilon}$, $d^{\text{GML}}(\boldsymbol{\varepsilon})$ is also the distance from $\mathbf{SAb}$ to the GML decision region of $\mathbf{SA}(\mathbf{b}-2\boldsymbol{\varepsilon})$. If the minimum $d^{\text{GML}}(\boldsymbol{\varepsilon})$ over all error vectors is equal to $d^{\text{GML}}(\mathbf{e}_k)$ where $\mathbf{e}_k$ is the $k$th coordinate vector, then by the GML detector the $k$th user achieves unit AME. Tse and Verdú proved [1] that the GML detector achieves almost surely unit AME in the LRS-CDMA. As an extension, the following theorem further indicates that the LRS-CDMA possess the LML characteristic. Let $E$ denote the set of error vectors and $I(\boldsymbol{\varepsilon})$ and $w(\boldsymbol{\varepsilon})$ the index set of nonzero elements and weight of $\boldsymbol{\varepsilon}$, respectively.

*Theorem 1*: In the LRS-CDMA with any $\alpha > 0$, (i) given any positive integers $M_1 < M_2$, $d^{\text{GML}}(\boldsymbol{\varepsilon}_1) < d^{\text{GML}}(\boldsymbol{\varepsilon}_2)$ a.s. $\forall \boldsymbol{\varepsilon}_1, \boldsymbol{\varepsilon}_2 \in E$ such that $I(\boldsymbol{\varepsilon}_1) \subset I(\boldsymbol{\varepsilon}_2)$ and $w(\boldsymbol{\varepsilon}_1) \leq M_1 < w(\boldsymbol{\varepsilon}_2) \leq M_2$; (ii) for any $\delta > 0$, there exists $M \geq 2$ such that $d^{\text{GML}}(\boldsymbol{\varepsilon}) > \delta$ a.s. $\forall \boldsymbol{\varepsilon} \in E$ with $w(\boldsymbol{\varepsilon}) \geq M$. □

The LML characteristic means that in the LRS CDMA limit the mapping of $\mathbf{SA}$ from $\{-1,1\}^K$ to $\{\mathbf{SAb} : \mathbf{b} \in \{-1,1\}^K\} \subset \mathbb{R}^N$ retains the local topology of $\{-1,1\}^K$ at any bit vector. Standing at the transmitted signal in the $\mathbf{r}$ space, one would typically see that the error signals with larger error weights are farther and all the error signals with the error weights tending to infinity are infinitely far. Since the GML decision is based on the nearest distance from $\mathbf{r}$ to a signal, the GML BER in the high SNR regime is dominated by the signals that have one bit error. This suggests that to achieve the GML detection one would, without the exhaustive search over the entire set $\{-1,1\}^K$, perform only an LML detection.

## B. Achievability of single-user performance by WSLAS

For the WSLAS detector, let $l^{\text{LML}}(\boldsymbol{\varepsilon})$ be the hyperplane that passes through the vertex of the LML point region of $\mathbf{SA}(\mathbf{b}-2\boldsymbol{\varepsilon})$ and is parallel to the optimum hyperplane $l^{\text{GML}}(\boldsymbol{\varepsilon})$. It is obtained in [6] that the distance from $\mathbf{SAb}$ to the hyperplane $l^{\text{LML}}(\boldsymbol{\varepsilon})$ is equal to

$$d^{\text{LML}}(\boldsymbol{\varepsilon}) = \frac{\boldsymbol{\varepsilon}^T (2\mathbf{H} - \mathbf{A}^2) \boldsymbol{\varepsilon}}{\sqrt{\boldsymbol{\varepsilon}^T \mathbf{H} \boldsymbol{\varepsilon}}}. \tag{5}$$

The distance from $\mathbf{SAb}$ to the LML point region of $\mathbf{SA}(\mathbf{b}-2\boldsymbol{\varepsilon})$ is lower bounded by $d^{\text{LML}}(\boldsymbol{\varepsilon})$. Though $d^{\text{LML}}(\boldsymbol{\varepsilon}) \leq d^{\text{GML}}(\boldsymbol{\varepsilon})$ for all indecomposable error vectors $\boldsymbol{\varepsilon}$, $d^{\text{LML}}(\boldsymbol{\varepsilon}) = d^{\text{GML}}(\boldsymbol{\varepsilon})$ if $w(\boldsymbol{\varepsilon}) = 1$. Thus, regarding discrimination of signals with one bit error, the WSLAS detector performs equally well as the GML detector does. The following theorem reveals a relationship between the WSLAS and the GML detectors.

*Theorem 2*: In the LRS-CDMA, (i) for any $\alpha > 0$, given any positive integer $M$, $d^{\text{LML}}(\boldsymbol{\varepsilon}) \to d^{\text{GML}}(\boldsymbol{\varepsilon})$ a.s. $\forall \boldsymbol{\varepsilon} \in E$ with $w(\boldsymbol{\varepsilon}) \leq M$; moreover, (ii) if $\alpha < \alpha^* \equiv \frac{1}{2} - 1/(4\ln 2)$, then for any $\delta > 0$ there exists $M \geq 2$ such that $d^{\text{LML}}(\boldsymbol{\varepsilon}) > \delta$ a.s. $\forall \boldsymbol{\varepsilon} \in E$ with $w(\boldsymbol{\varepsilon}) \geq M$. □

Different from the GML decision region, the LML point regions of different bit vectors may be overlapped with each other. Consequently, the local optimality of the WSLAS detectors cannot guarantee the global optimality since an observation $\mathbf{r}$ may be located in the region where a number of LML points coexist, most of which have lower likelihoods than the GML. However, the following two theorems indicate that in the regime of $\alpha < \alpha^*$ and vanishing noise power, the suboptimal WSLAS detector can achieve the GML.

*Theorem 3*: In the LRS-CDMA with $\alpha < \alpha^*$, the AME's of all the LML detectors converge a.s. to one. □

Since Tse and Verdú proved [1] that the GML detector achieves unit AME in LRS-CDMA, none of other suboptimal detectors has been found to achieve unit AME but the WSLAS detector indicated by Theorem 3. Like the GML detector [1], the WSLAS detector achieving unit AME is obtained by taking the zero noise limit first and then large system limit. However, the following theorem further indicates that in a looser condition where the noise power tends to zero and the system size tends to infinity at the same rate, an LML point is almost surely the GML point.

*Theorem 4*: In the LRS-CDMA where $\alpha < \alpha^*$ and $\sqrt{N}\sigma = c \in (0, \infty)$ fixed, an LML point is a.s. the GML point. □

In the regime of $\alpha < \alpha^*$ and vanishing noise, there is typically no LML point but the GML point. The likelihood function is sufficiently smooth. Thus, the WSLAS detector that in each step searches a higher likelihood in the neighborhood of a tentatively obtained vector can reach the GML point. It is practically interesting that the WSLAS detector can be implemented with a per-bit complex linear in the number of transmitted bits while the GML detector is NP-hard.

All the results are applicable to the LML detectors with neighborhood size $J \geq 2$, which are also LML detectors with neighborhood size one and can be implemented with a per-bit complexity of order $\binom{K}{J}$ [7].

### IV. SIMULATION RESULTS

Simulations are carried out for the QLRS-CDMA where all users have the identical $B_k = B$ and equal power. For $BK \leq 128$, long spreading sequences are used. For $BK > 128$, short sequences are used and the BER's for five samples of short sequences are estimated and shown together with their averages. Four GPLAS detectors [6] with group sizes $J = 8, 4, 2,$ and 1 are cascaded in the order of group size. That is, the output of the GPLAS detector with group size 8 is the initial of the GPLAS detector with group size 4. The cascaded GPLAS detectors from size 8 through 1 form a WSLAS detector. The GPLAS detectors cyclically update bits group by group and the SLAS detector updates bits cyclically bit by bit. The BER of the fixed point for each of the LAS detectors is shown. The bit flip rate (BFR) is the total number of bit flips divided by the total number of bits tested. The per-bit complexity, defined as the average number of additions per transmission divided by $BK$, is equal to BFR times $BK$ [6]. Thus, the BFR indicating the complexity is also shown.

As performance references, the BER's of the MMSE-DF with a per-bit complexity $1.5BK$ and the SIC detector [12] are estimated in simulation. The BER's of the MF, decorrelator, MMSE, and GML detectors in the LRS-CDMA limit are also shown. The first three are obtained by the $Q$ function evaluated at the limit SIR [8][9] since the interference of these linear receivers is asymptotically Gaussian [10] and the fourth is calculated by Tanaka's formula [11]. All these suboptimal detectors are linearly complex. In all simulations, only the multiplication $BK$ is given and thus the results are applicable to any pair of integers $B$ and $K$ with the given $BK$.

The BER's of the LAS detectors shown in Fig. 1 monotonically decrease as $BK$ increases, which justifies the proposal to construct the QLRS-CDMA. As $BK$ increases, the variance of crosscorrelations of the random sequences $1/(BN) = \alpha/(BK)$ decreases, the distances $2d^{GML}(\varepsilon)$ from the transmitted signal to the error signals are eventually dominated by their error weights and the channel eventually presents the LML characteristic. Consequently, the LAS detectors can perform better by exploiting the LML characteristic. The BER's of the GPLAS detectors with small group sizes approach the GML BER when $BK$ is greater than 500, confirming Theorems 3 and 4.

From the figure, we can see how the BER's of the GPLAS detectors monotonically decreases to the limit GML BER $2.1 \times 10^{-4}$ as $B$ increases with a fixed $K$. For example, consider a system of $K = 8$ users. The BER's averaged over the five sequence samples with $J = 2, 1$ are about $\{7.1 \times 10^{-2}, 6.5 \times 10^{-2}\}$ for $B = 1$ (no bit extending and multiplexing), $\{1.1 \times 10^{-2}, 5.4 \times 10^{-3}\}$ for $B = 16$, $\{2.9 \times 10^{-4}, 2.7 \times 10^{-4}\}$ for $B = 64$, $\{2.1 \times 10^{-4}, 2.0 \times 10^{-4}\}$ for $B = 256$, and $\{2.1 \times 10^{-4}, 2.1 \times 10^{-4}\}$ for $B = 416$. The BER's with other $K$'s can be also obtained from the figure.

Theorems 3 and 4 are verified with a much wider regime of SNR by the simulation result in Fig. 2 where the GPLAS detectors with group size 1, 2 approach the GML BER in *all* SNR and the single-user bound in high SNR. In contrast, all other suboptimal detectors present a vast gap from the single-user BER in high SNR. The complexity of the LAS detectors is slightly affected by SNR, and is less than $0.33BK$ when the initial detector is the MF.

Theorems 3 and 4 are verified in a much wider regime of $\alpha$ by the simulation result in Fig. 3. For $\alpha \leq 1.0$, the GPLAS detectors with $J = 1$ and 2 perform as well as the GML detector does, all close to the single-user bound. As $\alpha$ increases, the LAS detectors are slightly affected while other suboptimal detectors are apparently suffered from the increasing interference. For $\alpha > 1.0$, the BER's of the LAS detectors sharply increase around the transient load $\alpha_T = 1.1$ while the GML detector behaves similarly at 1.32 that takes on three BER's for $\alpha > 1.32$. In this sense, with a target BER close to the single-user bound, the spectral efficiency equals 1.1 for the GPLAS and 1.32 bits/s/Hz for the GML. The former is only 17% lower but the complexity to reach the former is linear and NP-hard to the latter. The BER variation of the LAS detectors is large for $\alpha > 1.1$, which suggests that the LAS detectors may also take multiple BER's in the limit LRS-CDMA for large $\alpha$. As shown in (b), the complexity of all LAS detectors is less than $0.5BK$ with the initial MF detector. Meanwhile, for $\alpha \leq 1$, the initial detector has no affect on the LAS BER when $BK$ is over 500, contradicting the common view that performance of an iterative detector significantly depends on its initial detector. However, the transient load $\alpha_T$ is sensitive to the initial detector, and the initial MF increases $\alpha_T$ by about 0.05 compared with the random initial (results are not shown).

In all the simulations, the SLAS detector approaches almost the same performance of the WSLAS detector since both are LML detectors. However, the former has a little higher complexity. The initial MF, which is almost cost-free, can significantly reduce the complexity of the LAS detectors. The per-bit complexity of the LAS detectors in all cases is about $cBK$ with $c \leq 0.5$. Hence, the computation cost incurred in the QLRS-CDMA increases only linearly with the extending factor $B$. Being also LML detectors with neighborhood size one and having higher complexity, the LML detectors with neighborhood size $J \geq 2$ [7] are expected to perform better than the WSLAS detector.

Though this work considers bit-synchronous uncoded data transmission over Gaussian CDMA channels, the QLRS-CDMA employing the LAS and LML detectors is obviously applicable to bit-asynchronous, coded data transmission, fading, multipath, MIMO, or/and multicarrier CDMA channels. The spreading in both time and frequency can combat dispersion both in frequency and time when the QLRS-CDMA is applied in fading channels.

## APPENDIX

*Proof of Theorem 1*:

The proof of (i) follows the similar lines in [1]. Consider $\varphi_K$ the event that $d^{\text{GML}}(\boldsymbol{\varepsilon}_1) \geq d^{\text{GML}}(\boldsymbol{\varepsilon}_2)$ for some $(\boldsymbol{\varepsilon}_1, \boldsymbol{\varepsilon}_2) \in G$ where $G$ is the set of pairs $(\boldsymbol{\varepsilon}', \boldsymbol{\varepsilon}'')$ of error vectors such that $I(\boldsymbol{\varepsilon}') \subset I(\boldsymbol{\varepsilon}'')$ and $w(\boldsymbol{\varepsilon}') \leq M_1 < w(\boldsymbol{\varepsilon}'') \leq M_2$. It shall be shown that $\Pr(\varphi_K)$ converges to zero exponentially as $K$ tends to infinity. Then result (i) follows from this together with the Borel-Cantelli lemma. By the union bound,

$$\Pr(\varphi_K) \leq \sum_{(\boldsymbol{\varepsilon}_1, \boldsymbol{\varepsilon}_2) \in G} \Pr\{[d^{\text{GML}}(\boldsymbol{\varepsilon}_1)]^2 - [d^{\text{GML}}(\boldsymbol{\varepsilon}_2)]^2 \geq 0\}. \tag{6}$$

The number of terms in $G$ is polynomial in $K$. Then it suffices that each probability in (6) converges to zero exponentially in $K$. Since $[d^{\text{GML}}(\boldsymbol{\varepsilon}_m)]^2 = (1/N)\sum_{j=1}^{N}\left(\sum_{i \in I(\boldsymbol{\varepsilon}_m)} s_{ji} A_i \varepsilon_{mi}\right)^2$ for $m = 1, 2$, it follows that $[d^{\text{GML}}(\boldsymbol{\varepsilon}_1)]^2 - [d^{\text{GML}}(\boldsymbol{\varepsilon}_2)]^2 = (1/N)\sum_{J=1}^{N} Z_j$ where $Z_j = -\left(\sum_{i \in I(\boldsymbol{\varepsilon}_2 - \boldsymbol{\varepsilon}_1)} s_{ji} A_i \varepsilon_{2i}\right)^2 - 2\sum_{i \in I(\boldsymbol{\varepsilon}_2 - \boldsymbol{\varepsilon}_1)} s_{ji} A_i \varepsilon_{2i} \sum_{l \in I(\boldsymbol{\varepsilon}_1)} s_{jl} A_l \varepsilon_{1l}$. $Z_j$'s are i.i.d. random variables with mean $E(Z_1) = -\sum_{i \in I(\boldsymbol{\varepsilon}_2 - \boldsymbol{\varepsilon}_1)} A_i^2$ which is less than zero. Thus, each event in the union bound is a large deviation. It can be obtained that $|Z_1| \leq w(\boldsymbol{\varepsilon}_2 - \boldsymbol{\varepsilon}_1)(w(\boldsymbol{\varepsilon}_2 - \boldsymbol{\varepsilon}_1) + 2w(\boldsymbol{\varepsilon}_1))A''^2$ which is finite and therefore $E[\exp(a|Z_1|)] < \infty$ for any fixed $a > 0$. By the lemma of large deviations [13] (pp. 281), there exists $c \in [0, a]$ such that for each $\delta > 0$,

$$\Pr\{[d^{\text{GML}}(\boldsymbol{\varepsilon}_1)]^2 - [d^{\text{GML}}(\boldsymbol{\varepsilon}_2)]^2 > E(Z_1) + \delta\} \leq g^N(c), \tag{7}$$

$$g(c) \equiv \exp[-c(E(Z_1) + \delta)]E[\exp(cZ_1)] < 1. \tag{8}$$

Now, it is sufficient to find $\delta > 0$ such that $E(Z_1) + \delta \leq 0$ uniformly for all $A_i$'s and all $(\boldsymbol{\varepsilon}', \boldsymbol{\varepsilon}'') \in G$. Since $E(Z_1) \leq -A'^2$ uniformly, then take $\delta = A'^2$. This completes proof of (i). Result (ii) is proved by Step 2 of [1] with the change that $\mathbf{v}^T\mathbf{ARAv} < 1$ on page 2721 of [1] is replaced by $\mathbf{v}^T\mathbf{ARAv} < \delta$ and some lines are modified accordingly. □

*Proof of Theorem 2:*

(i) Consider $\varphi_K$ the event that for a fixed $\delta > 0$, $|d^{\text{LML}}(\boldsymbol{\varepsilon}) - d^{\text{GML}}(\boldsymbol{\varepsilon})| > \delta$ for some $\boldsymbol{\varepsilon} \in E$ with $w(\boldsymbol{\varepsilon}) \leq M$. It shall be shown that $\Pr(\varphi_K)$ converges to zero exponentially as $N$ tends to infinity. Then result (i) follows from this together with the Borel-Cantelli lemma. By the union bound,

$$\Pr(\varphi_K) \leq \sum_{w(\boldsymbol{\varepsilon}) \leq M} \Pr\left[|d^{\text{LML}}(\boldsymbol{\varepsilon}) - d^{\text{GML}}(\boldsymbol{\varepsilon})| > \delta\right]. \tag{9}$$

The number of error vectors with $1 \leq w(\boldsymbol{\varepsilon}) \leq M$ is polynomial in $K$. Then it suffices that each probability in (9) converges to zero exponentially in $K$.

For each $0 < \mu < A'$,

$$\Pr\left[|d^{\text{LML}}(\mathbf{\epsilon}) - d^{\text{GML}}(\mathbf{\epsilon})| > \delta\right] = \Pr\left(\left|\sqrt{\mathbf{\epsilon}^T \mathbf{H}\mathbf{\epsilon}} - \|\mathbf{A}\mathbf{\epsilon}\|^2 / \sqrt{\mathbf{\epsilon}^T \mathbf{H}\mathbf{\epsilon}}\right| > \delta\right)$$

$$= \Pr\left(\left|\sqrt{\mathbf{\epsilon}^T \mathbf{H}\mathbf{\epsilon}} - \|\mathbf{A}\mathbf{\epsilon}\|^2 / \sqrt{\mathbf{\epsilon}^T \mathbf{H}\mathbf{\epsilon}}\right| > \delta; \mathbf{\epsilon}^T \mathbf{H}\mathbf{\epsilon} \geq \mu\right) + \Pr\left(\left|\sqrt{\mathbf{\epsilon}^T \mathbf{H}\mathbf{\epsilon}} - \|\mathbf{A}\mathbf{\epsilon}\|^2 / \sqrt{\mathbf{\epsilon}^T \mathbf{H}\mathbf{\epsilon}}\right| > \delta; \mathbf{\epsilon}^T \mathbf{H}\mathbf{\epsilon} < \mu\right)$$

$$\leq \Pr\left(\left|\mathbf{\epsilon}^T \mathbf{H}\mathbf{\epsilon} - \|\mathbf{A}\mathbf{\epsilon}\|^2\right| > \delta\sqrt{\mathbf{\epsilon}^T \mathbf{H}\mathbf{\epsilon}}\right) + \Pr\left(\mathbf{\epsilon}^T \mathbf{H}\mathbf{\epsilon} < \mu\right). \tag{10}$$

It has been shown in [1] that the second probability converges to zero exponentially and uniformly for all $A_i$'s and all $\mathbf{\epsilon}$ with $w(\mathbf{\epsilon}) \leq M$.

It is sufficient to show in what follows that the first probability in (10) also converges to zero exponentially and uniformly for all $A_i$'s and all $\mathbf{\epsilon}$ with $w(\mathbf{\epsilon}) \leq M$. To this end, define

$$Y_i = \sum_{j \in I(\mathbf{\epsilon})} s_{ij} A_j \varepsilon_j / \|\mathbf{A}\mathbf{\epsilon}\|, \tag{11}$$

which are i.i.d. with zero mean and unit variance and then

$$Y = \sum_{i=1}^{N} Y_i^2 \tag{12}$$

has mean $N$. Since $\mathbf{\epsilon}^T \mathbf{H}\mathbf{\epsilon} = \|\mathbf{A}\mathbf{\epsilon}\|^2 Y/N$, the first probability in (10) equals

$$\Pr\left(\left|\mathbf{\epsilon}^T \mathbf{H}\mathbf{\epsilon} - \|\mathbf{A}\mathbf{\epsilon}\|^2\right| > \delta\sqrt{\mathbf{\epsilon}^T \mathbf{H}\mathbf{\epsilon}}\right) = \Pr\left(\|\mathbf{A}\mathbf{\epsilon}\||Y - N| > \delta\sqrt{NY}\right)$$

$$= \Pr\left(\|\mathbf{A}\mathbf{\epsilon}\|(Y - N) > \delta\sqrt{NY}\right) + \Pr\left(\|\mathbf{A}\mathbf{\epsilon}\|(Y - N) < -\delta\sqrt{NY}\right). \tag{13}$$

By simple algebra, the first probability in (13) equals

$$\Pr\left(\|\mathbf{A}\mathbf{\epsilon}\|(Y - N) > \delta\sqrt{NY}\right) = \Pr(Y > \beta N), \tag{14}$$

$$\beta = \left(\delta + \sqrt{\delta^2 + 4\|\mathbf{A}\mathbf{\epsilon}\|^2}\right)^2 / (4\|\mathbf{A}\mathbf{\epsilon}\|^2) \tag{15}$$

is greater than one for all $\delta > 0$. Note that the variance of $Y_1$ equals one and thus the event in the considered probability is a large deviation. Since $Y_1^2 \leq w(\mathbf{\epsilon})A''^2/A'^2$, $E[\exp(aY_1^2)] < \infty$ for any fixed $a > 0$. It follows from the lemma of large deviations [13] that there exists $c \geq 0$ such that for each $\delta > 0$,

$$\Pr\left(\|\mathbf{A}\mathbf{\epsilon}\|(Y - N) > \delta\sqrt{NY}\right) \leq g^N(c, \beta), \tag{16}$$

$$g(c, \beta) \equiv \exp(-c\beta)E[\exp(cY_1^2)] < 1. \tag{17}$$

For each $\delta > 0$, $\beta > 1$ is true uniformly for all $A_i$'s and all $\mathbf{\epsilon}$ with $w(\mathbf{\epsilon}) \leq M$. Hence, the probability (16) converges to zero exponentially and uniformly.

Similarly, the second probability in (13) equals

$$\Pr\left(\|\mathbf{A}\mathbf{\epsilon}\|(Y - N) < -\delta\sqrt{NY}\right) = \Pr(Y < \beta^{-1}N). \tag{18}$$

Note that for all $\delta > 0$, $\beta^{-1}$ is less than one while the variance of $Y_1$ equals one. Since $E[\exp(aY_1^2)] < \infty$ for any fixed $a < 0$, in terms of the lemma of large deviations [13] there exists $c \leq 0$ such that for each $\delta > 0$,

$$\Pr\left(\|\mathbf{A}\boldsymbol{\varepsilon}\|(Y-N) < -\delta\sqrt{NY}\right) \leq g^N(c,\beta^{-1}) \tag{19}$$

and $g(c, \beta^{-1}) < 1$. For each $\delta > 0$, $\beta^{-1} < 1$ is true uniformly. Hence, the probability (18) converges to zero exponentially and uniformly. This completes the proof of (i).

(ii) Let $F$ be the set of indecomposable error vectors. It has been shown in [6] that for each $\boldsymbol{\varepsilon} \in E$, there is $\boldsymbol{\varepsilon}' \in F$ such that $d^{\text{LML}}(\boldsymbol{\varepsilon}') < d^{\text{LML}}(\boldsymbol{\varepsilon})$. Then it is sufficient to show that the result is true for all indecomposable error vectors $\boldsymbol{\varepsilon} \in F$. Let $G_K$ be the event that given $\delta > 0$, $d^{\text{LML}}(\boldsymbol{\varepsilon}) < \delta$ for some $\boldsymbol{\varepsilon} \in F$ with $w(\boldsymbol{\varepsilon}) \geq M$. It shall be shown that $\Pr(G_K)$ converges to zero exponentially fast as $N$ tends to zero. Together with the Borel-Cantelli lemma, it proves the result. The probability of $G_K$ is upper bounded by

$$\Pr(G_K) \leq \sum_{\boldsymbol{\varepsilon} \in E, w(\boldsymbol{\varepsilon}) \geq M} \Pr\left[d^{\text{LML}}(\boldsymbol{\varepsilon}) \leq \delta; \boldsymbol{\varepsilon} \in F\right] = \sum_{m=M}^{K} \sum_{|J|=m} \sum_{\boldsymbol{\varepsilon} \in \psi(J)} \Pr\left[d^{\text{LML}}(\boldsymbol{\varepsilon}) \leq \delta; \boldsymbol{\varepsilon} \in F\right]. \tag{20}$$

Since by Lemma 1 of [6] there are at most two indecomposable error vectors in $\psi(J)$ regardless of $\mathbf{S}$ and $\mathbf{A}$,

$$\Pr(G_K) \leq 2\sum_{m=M}^{K} \sum_{|J|=m} \Pr(d^{\text{LML}}(\boldsymbol{\varepsilon}) \leq \delta; \boldsymbol{\varepsilon} \in F; I(\boldsymbol{\varepsilon}) = J) \leq 2\sum_{m=M}^{K} \sum_{|J|=m} \Pr(d^{\text{LML}}(\boldsymbol{\varepsilon}) \leq \delta; I(\boldsymbol{\varepsilon}) = J) \tag{21}$$

where $\boldsymbol{\varepsilon}$ is any error vector in $\psi(J)$ with given $J$. Note that

$$d^{\text{LML}}(\boldsymbol{\varepsilon}) = \|\mathbf{A}\boldsymbol{\varepsilon}\|\left(2\sqrt{Y/N} - \sqrt{N/Y}\right). \tag{22}$$

To get rid of the event $Y = 0$, consider the probability conditioned with fixed $\mu \in (0, \frac{1}{2})$ and then by simple algebra

$$\Pr\left(d^{\text{LML}}(\boldsymbol{\varepsilon}) \leq \delta \mid Y > \mu\right) = \Pr\left(Y \leq N\lambda \mid Y > \mu\right), \tag{23}$$

$$\lambda = (\delta + \sqrt{\delta^2 + 8\|\mathbf{A}\boldsymbol{\varepsilon}\|^2})^2 / (16\|\mathbf{A}\boldsymbol{\varepsilon}\|^2), \tag{24}$$

which is greater than $\frac{1}{2}$ and converges to $\frac{1}{2}$ as $M \to \infty$. Thus for a fixed $J$ and any $\boldsymbol{\varepsilon} \in \psi(J)$,

$$\Pr\left(d^{\text{LML}}(\boldsymbol{\varepsilon}) \leq \delta\right) = \Pr\left(d^{\text{LML}}(\boldsymbol{\varepsilon}) \leq \delta \mid Y > \mu\right)\Pr(Y > \mu) + \Pr\left(d^{\text{LML}}(\boldsymbol{\varepsilon}) \leq \delta \mid Y \leq \mu\right)\Pr(Y \leq \mu)$$

$$\leq \Pr\left(Y \leq N\lambda \mid Y > \mu\right)\Pr(Y > \mu) + \Pr(Y \leq \mu) = \Pr(Y \leq N\lambda) \tag{25}$$

where the last equality holds since $N\lambda \geq \frac{1}{2}$.

It follows from the Chernoff bound that $\Pr(Y \leq N\lambda) \leq \exp(-t\lambda N)\{E[\exp(tY_1^2)]\}^N$ for any $t < 0$. If $s_{ji}$'s were i.i.d. Gaussian with zero mean and unit variance, then $Y_1^2$ would be Chi-square distributed with degree one. An upper bound on the difference between the moment-generating functions of $Y_1^2$ and a Chi-square random variable has been given in [1] that $\left|E[\exp(tY_1^2)] - E[\exp(tZ^2)]\right| < a/\sqrt{w(\boldsymbol{\varepsilon})}$ for some constant $a$ where $Z \sim N(0,1)$. The moment-generating function for the Chi-square distribution is $E[\exp(tZ^2)] = 1/\sqrt{1-2t}$. Then

$$\Pr\left(d^{\text{LML}}(\boldsymbol{\varepsilon}) \leq \delta\right) < \exp(-t\lambda N)\left\{E[\exp(tZ^2)] + a/\sqrt{w(\boldsymbol{\varepsilon})}\right\}^N = \exp(-t\lambda N)\left(1/\sqrt{1-2t} + a/\sqrt{w(\boldsymbol{\varepsilon})}\right)^N$$

$$= \exp\left[-N\left(t\lambda - \ln\left(1/\sqrt{1-2t} + a/\sqrt{w(\boldsymbol{\varepsilon})}\right)\right)\right] \tag{26}$$

where $t\lambda - \ln\left(1/\sqrt{1-2t} + a/\sqrt{w(\varepsilon)}\right)$ converges to $t/2 + (1/2)\ln(1-2t)$ as $M$ increases. It follows from (21) that for sufficiently large $M$

$$\Pr(G_K) < 2\sum_{m=M}^{K}\sum_{|J|=m}\exp\left[-N\left((1/2)t + (1/2)\ln(1-2t)\right)\right] < 2^{K+1}\exp\left[-N\left((1/2)t + (1/2)\ln(1-2t)\right)\right]$$

$$= 2\exp\left[-N\left((1/2)t - \alpha\ln 2 + (1/2)\ln(1-2t)\right)\right]. \tag{27}$$

where the exponent is minimized by taking $t = -\frac{1}{2}$ and then (27) converges to zero exponentially if $\alpha < \alpha^*$. This completes the proof of Theorem 2. □

*Proof of Theorem 3:*

It is obtained in [6] that the AME of an LML detector is lower bounded by the least $[d^{\text{LML}}(\varepsilon)]^+/A_k$ for all $\varepsilon \in F_k$ where $F_k$ is the set of indecomposable error vectors affecting user $k$ and $[x]^+ = \max\{0, x\}$. For $\varepsilon$ with $w(\varepsilon) = 1$, it follows from (5) that $[d^{\text{LML}}(\varepsilon)]^+/A_k = 1$. For all $\varepsilon$ with $2 \le w(\varepsilon) \le M$ where $M \ge 2$ is fixed, $d^{\text{LML}}(\varepsilon) \to d^{\text{GML}}(\varepsilon)$ almost surely by Theorem 2 and then $d^{\text{LML}}(\varepsilon)/A_k > 1$ almost surely by Theorem 1. By means of Theorem 2, there exists an $M$ such that $d^{\text{LML}}(\varepsilon)/A_k > 1$ almost surely for all $\varepsilon$ with $w(\varepsilon) > M$. Since AME cannot be greater than one, the AME of all LML/WSLAS detectors converges almost surely to one. □

*Proof of Theorem 4:*

Given an observation $\mathbf{r} \in \mathbb{R}^N$, let $\mathbf{b}^{\text{LML}}$ be an LML vector defined with neighborhood size one. Denote $\varepsilon = \frac{1}{2}(\mathbf{b} - \mathbf{b}^{\text{LML}})$ and consider a fixed positive integer $M$.

First, consider all error vectors with $w(\varepsilon) = 1$. Specifically, let $\mathbf{b}_i$ be the vector differing from $\mathbf{b}^{\text{LML}}$ by the $i$th bit. Since $\mathbf{b}_i$ is in the neighborhood of $\mathbf{b}^{\text{LML}}$, it satisfies

$$\|\mathbf{r} - \mathbf{S}\mathbf{A}\mathbf{b}_i\|^2 - \|\mathbf{r} - \mathbf{S}\mathbf{A}\mathbf{b}^{\text{LML}}\|^2 = -2(\mathbf{r} - \mathbf{S}\mathbf{A}\mathbf{b}^{\text{LML}})^T\mathbf{S}\mathbf{A}(\mathbf{b}_i - \mathbf{b}^{\text{LML}}) + \|\mathbf{A}(\mathbf{b}_i - \mathbf{b}^{\text{LML}})\|^2 \ge 0 \tag{28}$$

Second, consider all error vectors $\mathbf{b}$ such that $2 \le w(\varepsilon) \le M$. Let $G_K$ be the event that there are some $\mathbf{b}$ such that $\|\mathbf{r} - \mathbf{S}\mathbf{A}\mathbf{b}\|^2 - \|\mathbf{r} - \mathbf{S}\mathbf{A}\mathbf{b}^{\text{LML}}\|^2 < -\delta$ for $\delta > 0$. Then

$$\Pr(G_K) \le \sum_{2 \le w(\varepsilon) \le M} \Pr(\|\mathbf{r} - \mathbf{S}\mathbf{A}\mathbf{b}\|^2 - \|\mathbf{r} - \mathbf{S}\mathbf{A}\mathbf{b}^{\text{LML}}\|^2 < -\delta). \tag{29}$$

For each $\mathbf{b}$, we have

$$\|\mathbf{r} - \mathbf{S}\mathbf{A}\mathbf{b}\|^2 - \|\mathbf{r} - \mathbf{S}\mathbf{A}\mathbf{b}^{\text{LML}}\|^2 = -2(\mathbf{r} - \mathbf{S}\mathbf{A}\mathbf{b}^{\text{LML}})^T\mathbf{S}\mathbf{A}(\mathbf{b} - \mathbf{b}^{\text{LML}}) + \|\mathbf{S}\mathbf{A}(\mathbf{b} - \mathbf{b}^{\text{LML}})\|^2$$

$$= -\sum_{i \in I(\varepsilon)} 2(\mathbf{r} - \mathbf{S}\mathbf{A}\mathbf{b}^{\text{LML}})^T\mathbf{S}\mathbf{A}(\mathbf{b}_i - \mathbf{b}^{\text{LML}}) + \|\mathbf{S}\mathbf{A}(\mathbf{b} - \mathbf{b}^{\text{LML}})\|^2. \tag{30}$$

Due to (28), $\Pr(G_K) \le \sum_{2 \le w(\varepsilon) \le M}\Pr\left(\|\mathbf{S}\mathbf{A}(\mathbf{b} - \mathbf{b}^{\text{LML}})\|^2 - \sum_{i \in I(\varepsilon)}\|\mathbf{A}(\mathbf{b}_i - \mathbf{b}^{\text{LML}})\|^2 < -\delta\right)$. Since $\|\mathbf{S}\mathbf{A}(\mathbf{b} - \mathbf{b}^{\text{LML}})\|^2 = (1/N)\sum_{j=1}^{N}Z_j^2$ where $Z_j = 2\sum_{i \in I(\varepsilon)}s_{ji}A_i\varepsilon_i$ and $E[Z_j^2] = \sum_{i \in I(\varepsilon)}\|\mathbf{A}(\mathbf{b}_i - \mathbf{b}^{\text{LML}})\|^2$. Hence, for each $\delta > 0$, $E_K(\mathbf{b}) = \{\|\mathbf{S}\mathbf{A}(\mathbf{b} - \mathbf{b}^{\text{LML}})\|^2 - E[Z_j^2] < -\delta\}$ is an event of large deviation. Since $|Z_1| \le 2A''w(\varepsilon)$, by the lemma of large

deviations [13] $\Pr[E_K(\mathbf{b})]$ converges to zero exponentially fast and so does $\Pr[G_K(\mathbf{b})]$ since the number of $\mathbf{b}$ considered is polynomial in $K$. By the Borel-Cantelli lemma, $\|\mathbf{r} - \mathbf{SAb}\|^2 \geq \|\mathbf{r} - \mathbf{SAb}^{LML}\|^2$ almost surely for all $\mathbf{b}$ such that $2 \leq w(\mathbf{\varepsilon}) \leq M$.

Third, consider all error vectors $\mathbf{b}$ such that $w(\mathbf{\varepsilon}) > M$. By the law of large numbers, $\mathbf{r}$ is almost surely located on the sphere of $\|\mathbf{r} - \mathbf{SAb}_0\| = c$ centered at the transmitted signal $\mathbf{SAb}_0$ with the finite radius $c$. Then $\|\mathbf{SA}(\mathbf{b}_0 - \mathbf{b}^{LML})\|$ is almost surely finite; otherwise, by Theorem 2 (ii) the distance from $\mathbf{SAb}_0$ to the LML region of $\mathbf{SAb}^{LML}$ would almost surely increase without bound and then $\mathbf{b}^{LML}$ would not be an LML point almost surely. Now we consider an $M$ such that the distance from $\mathbf{SAb}^{LML}$ to the LML region of $\mathbf{b}$ for all $\mathbf{b}$ with $w(\mathbf{\varepsilon}) \geq M$ is almost surely greater than $\gamma + 2c$ with a fixed $\gamma > 0$. By Theorem 2 (ii), such an $M$ exists. Then $\mathbf{r}$ is not located in the LML region of $\mathbf{b}$ and so $\mathbf{b}$ is not an LML point almost surely for all the $\mathbf{b}$'s with $w(\mathbf{\varepsilon}) \geq M$. This completes the proof. □

## ACKNOWLEDGEMENT

The author would like to thank the anonymous reviewers for their reviews that improve the quality of this letter.

## REFERENCES

[1] D. N. C. Tse and S. Verdú, "Optimum asymptotic multiuser efficiency of randomly spread CDMA," *IEEE Trans. Inform. Theory*, vol. 46, pp. 2718-2722, Nov. 2000.

[2] Y. Kabashima, "CDMA multiuser detection algorithm on the basis of belief propagation," *J. Phys. A: Math. Gen.*, vol. 36, pp. 11111-11121, Oct. 2003.

[3] T. Tanaka and M. Okada, "Approximate belief propagation, density evolution, and statistical neurodynamics for CDMA multiuser detection," *IEEE Trans. Inf. Theory*, vol. 51, no. 2, pp. 700-706, Feb. 2005.

[4] P. H. Tan and L. K. Rasmussen, "Belief propagation for coded multiuser detection," in *Proc. IEEE Int. Symp. Inform. Theory*, Seattle, Washington, pp. 1919-1923, July 2006.

[5] P. H. Tan and L. K. Rasmussen, "Asymptotically optimal nonlinear MMSE multiuser detection based on multivariate Gaussian approximation," *IEEE Trans. Commun.*, vol. 54, pp. 1427-1438, Aug. 2006.

[6] Y. Sun, "A family of likelihood ascent search multiuser detectors: an upper bound of bit error rate and a lower bound of asymptotic multiuser efficiency," submitted to *IEEE Trans. Commun*.

[7] Y. Sun, "Local maximum likelihood multiuser detection," in *Proc. 34th Ann. Conf. on Inform. Sci. and Syst.*, CISS'2001, pp. 7-12, The Johns Hopkins University, Baltimore, Maryland, March 21-23, 2001.

[8] D. N. C. Tse and S. V. Hanly, "Linear multiuser receivers: effective interference, effective bandwidth and user capacity," *IEEE Trans. Inform. Theory*, vol. 45, pp. 641-657, Mar. 1999.

[9] Y. C. Eldar and A. M. Chan, "On the asymptotic performance of the decorrelator," *IEEE Trans Inf. Theory*, vol. 49, no. 9, pp. 2309-2313, Sept. 2003.

[10] D. Guo, S. Verdú, and L. K. Rasmussen, "Asymptotic normality of linear multiuser receiver outputs," *IEEE Trans. Inform. Theory*, vol. 48, no. 12, pp. 3080- 3095, Dec. 2002.

[11] T. Tanaka, "A statistical-mechanics approach to large-system analysis of CDMA multiuser detectors," *IEEE Trans. on Inform. Theory*, vol. 48, pp. 2888-2910, Nov. 2002.

[12] S. Verdú, *Multiuser Detection*, Cambridge University Press, New York, 1998.

[13] B. Fristedt and L. Gray, *A Modern Approach to Probability Theory*. Birkhäuser, Boston, 1997.

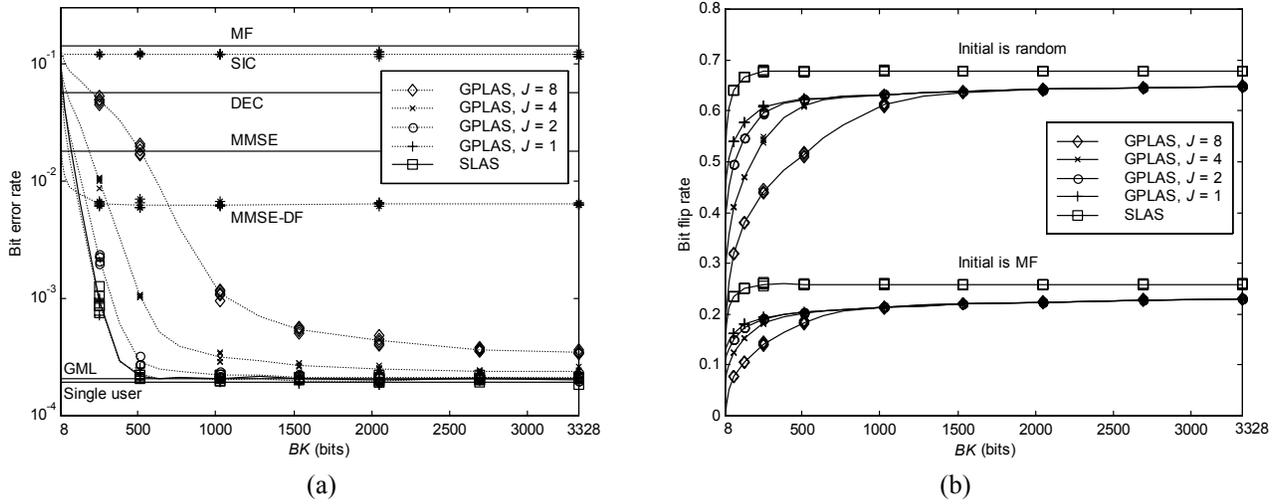

Fig. 1. (a) BER and (b) BFR versus *BK*. $\alpha = 0.8$, SNR = 11 dB and MF initial in (a).

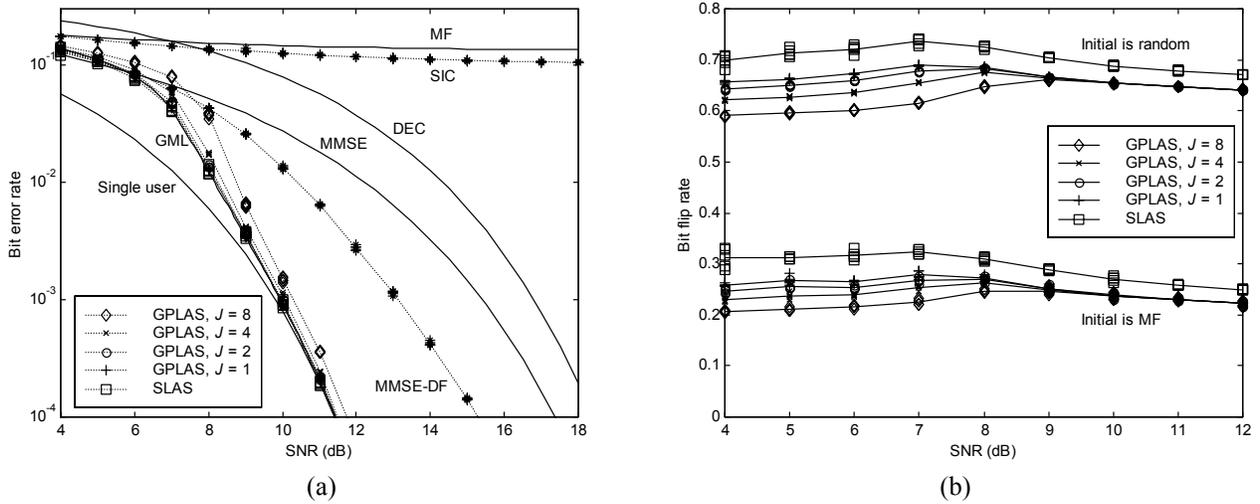

Fig. 2. (a) BER and (b) BFR versus SNR. $\alpha = 0.8$, *BK* = 3000 and random initial in (a).

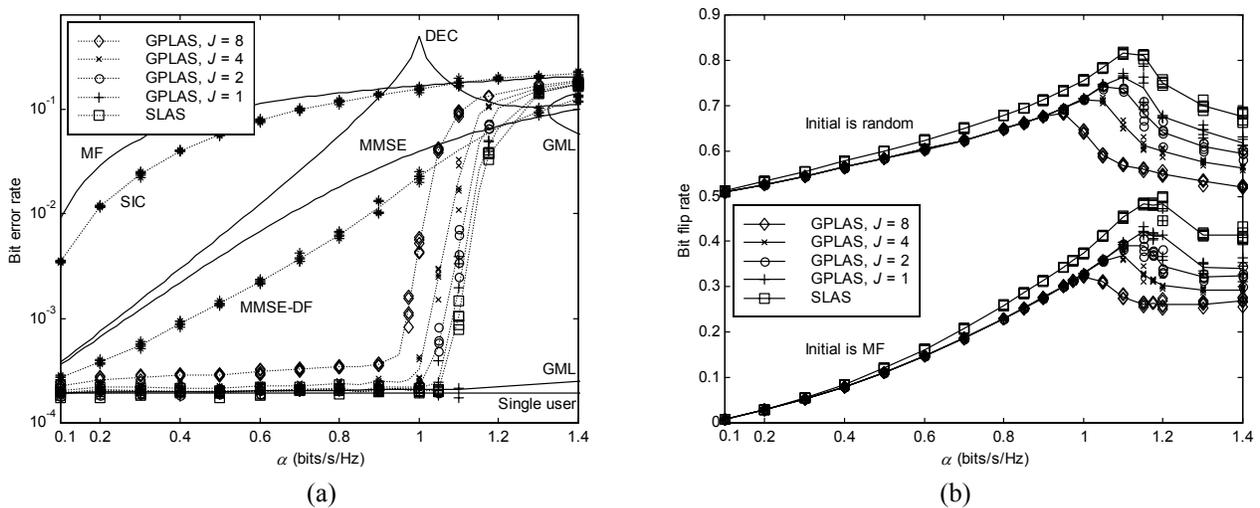

Fig. 3. (a) BER and (b) BFR versus $\alpha$. *BK* = 1136, 1600, 1960, 2264, 2536, 2784, 3000, and 3328 when $\alpha$ = 0.1, 0.2, 0.3, 0.4, 0.5, 0.6, 0.7, and $\alpha \geq 0.8$, respectively, SNR = 11 dB and MF initial in (a).